\documentclass[fleqn,10pt]{wlscirep}
\usepackage{subfig}
\usepackage{pdfpages}
\usepackage{graphicx}
\usepackage{epstopdf}
\usepackage{subfig}

\title{Atomic vapor as a source of tunable, non-Gaussian self-reconstructing optical modes}

\author[1]{Jon D. Swaim}
\author[1]{Kaitlyn N. David}
\author[1]{Erin M. Knutson}
\author[1]{Christian Rios}
\author[1]{Onur Danaci}
\author[1,*]{Ryan T. Glasser}
\affil[1]{Department of Physics, Tulane University, New Orleans, LA USA 70118}

\affil[*]{rglasser@tulane.edu}


\begin{abstract}
In this manuscript, we demonstrate the ability of nonlinear light-atom interactions to produce tunably non-Gaussian, partially self-healing optical modes.  Gaussian spatial-mode light tuned near to the atomic resonances in hot rubidium vapor is shown to result in non-Gaussian output mode structures that may be controlled by varying either the input beam power or the temperature of the atomic vapor.  We show that the output modes exhibit a degree of self-reconstruction after encountering an obstruction in the beam path.  The resultant modes are similar to truncated Bessel-Gauss modes that exhibit the ability to self-reconstruct earlier upon propagation than Gaussian modes. The ability to generate tunable, self-reconstructing beams has potential applications to a variety of imaging and communication scenarios. 
\end{abstract}
\begin{document}
\bibliographystyle{naturemag}

\flushbottom
\maketitle

\thispagestyle{empty}

\section*{Introduction}

Coherent light tuned on, or near, atomic resonances has allowed for the discovery and investigation of a wealth of nonlinear optical effects~\cite{Boyd2003, Gordon1965, Callen1967, Durbin1981, Boshier1982, LeBerre1984, Santamato1984, Deng2005, Nascimento2006}.  In addition to cross-action effects in which the presence of one beam alters the behavior of another, many self-action effects have been thoroughly investigated, including self-phase modulation~\cite{Durbin1981}, self-focusing~\cite{Shen1975, Marburger1975, Boshier1982} and self-defocusing~\cite{Gordon1965, Callen1967, Yu1998}, continuous-wave (CW) on-resonance enhancement~\cite{LeBerre1984}, and self-induced transparency~\cite{McCall1967, McCall1969}.  While nonlinear index-induced focusing and defocusing of light have been studied in both cold atom systems~\cite{Labeyrie2003, Labeyrie2007, Labeyrie2011} and hot atomic vapor~\cite{Grischkowsky1970, Bjorkholm1974, Zhang2015, Zhang2017}, these effects have only been studied individually, and to our knowledge no experiment to date has demonstrated a tunable interplay between nonlinear focusing and defocusing at a fixed detuning from an atomic resonance, with an input beam that is both converging and diverging through the nonlinear medium. Self-action effects strongly modify the spatial structure of an optical beam, and may result in output mode profiles that exhibit interesting optical properties such as diffraction-free and self-healing behavior~\cite{Durnin1987, Gori1987, MacDonald1996}.  The ability of optical modes to reconstruct themselves after encountering an obstruction in their path has been investigated in both the classical~\cite{Bouchal1998, Litvin2009, Chu2012, Wen2015} and quantum regimes~\cite{McLaren2014}, the latter of which has resulted in a demonstration of more robust propagation and recovery of quantum entanglement.  

The first experimental investigation into self-healing Bessel-Gauss (BG) beams was carried out by Durnin and co-authors~\cite{Durnin1987}.  In their work, significant (annular) aperturing of the incident optical mode produced a BG beam with reduced diffractive spreading compared to that of a Gaussian beam.  More recent methods involve either spatial light modulators~\cite{Davis1993, Vaity2015} or axicon (conical) lenses~\cite{McLeod1954, Bouchal1998}.  In some cases, spatial light modulators can be damaged at relatively low intensities~\cite{Beck2010, HamamatsuSLMTechInfo}, thus limiting the intensity of the generated non-Gaussian light. Further, while some progress has been made combining numerous optical elements with axicon lenses to produce tunable output self-healing modes, the cone angle of axicon lenses themselves is inherently fixed.  In contrast, hot atomic vapors provide an all-optical method of generating non-Gaussian, BG-like modes~\cite{Nascimento2006}.  With this method, two recent motivating works demonstrated some tunability in the shape of the output~\cite{Zhang2015, Zhang2017}.  While self-focusing and defocusing were the nonlinear effects considered in the theoretical discussions therein, a description of the spatial-mode structure of these modes, as well as their ability to self-reconstruct after encountering an obstruction still remains to be investigated.  

Here we experimentally demonstrate tunable generation of partially self-reconstructing non-Gaussian beams.  To accomplish this, we focus strong, Gaussian spatial-mode, nearly resonant laser light into a nonlinear medium consisting of hot alkali atoms.  A complex interplay between various self-action effects in the medium determines the overall output mode shape, and accordingly enables some tunability of the generated non-Gaussian beam by varying the input optical power or the temperature of the atomic medium.  The ability to tune the mode conversion process in this manner should allow for optimization of the self-reconstruction, which we demonstrate for a non-Gaussian beam encountering an obstruction at its center.  These findings suggest that tunable, non-Gaussian light generated via atomic vapors may be useful for experiments in quantum information, as well as demonstrating new approaches to optical communication and imaging.

\begin{figure}[ht!]
\centering
\includegraphics*[width=\textwidth]{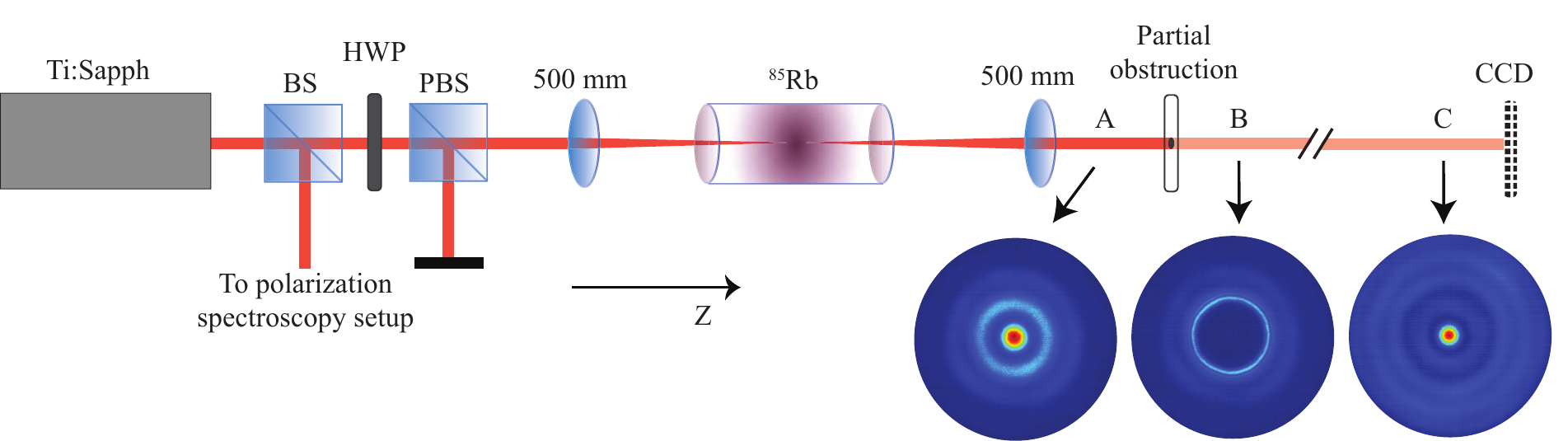}
\caption{\textbf{Schematic of the experimental setup for the generation of tunable, self-reconstructing optical modes}.  The mode images (A, B, and C) were obtained with an input power of $P = 300$ mW and a cell temperature of $T = 125$  $^\circ$C.  Ti:Sapph: Titanium-sapphire CW laser.  BS: Beamsplitter.  HWP: Half-wave plate.  PBS: Polarizing beam splitter. $^{85}$Rb:  Rubidium-85 vapor cell.  CCD: Camera.  A: Reference image before the obstruction.  B: Obstructed mode directly after the obstruction.  C: Reconstructed mode several meters after the obstruction.} 
\label{fig:setup}
\end{figure}

\section*{Results}

\subsection*{Experimental layout}
The experimental layout is shown in Fig.~\ref{fig:setup}.  Light from a CW titanium-sapphire laser is tuned close to the D1 line of rubidium and coupled into a single-mode, polarization-maintaining fiber in order to produce a Gaussian spatial mode.  The beam is focused to a spot size of 220\,$\mu$m at the center of a 2.5\,cm long rubidium-85 cell, whose temperature is controlled by a resistive heating element and monitored by a thermocouple.  A small portion of the light is sent to a polarization spectroscopy setup for laser locking (see Methods), and the remaining optical power is varied by the combination of a half-wave plate and polarizing beam splitter prior to the rubidium cell.  At a distance 190 mm after the cell, the resulting output modes are imaged with a CCD camera (2048 x 1088 pixels with a 5.5 $\mu$m pixel pitch) to investigate the spatial mode profiles as a function of input power and cell temperature.  In the reconstruction experiments, the output beam is instead re-collimated with a 500\,mm focal length lens placed one focal length from the center of the rubidium cell.  Additionally, due to the large size of the optical modes, a 60 mm focal length lens is placed approximately 30 mm in front of the camera.  Because atomic vapor acts as a nonlinear lens in the conversion process, the defocusing strength of the output mode varies depending on the choice of detuning, input power and cell temperature. While this has the benefit of producing the desired tunability of the mode shapes,  for consistency throughout all experiments the 500\,mm lens is kept fixed at one focal length from the focus of the unconverted Gaussian beam.  A removable circular obstruction 3 mm in diameter is placed in the center of the beam's path, and both the unobstructed and reconstructed images are recorded at various distances from the obstruction.  

\subsection*{Tunable non-Gaussian beam generation}
A typical output spatial mode is shown at position A in Fig.~\ref{fig:setup}.   These truncated BG-like modes are generated when the laser is detuned to the red side of the rubidium D1 line, over a frequency range on the order of one Doppler-broadened atomic linewidth.  In the medium, the Kerr nonlinearity results in a spatially-varying refractive index~\cite{Boyd2003}, which in turn leads to a spatially-varying phase shift on the incident beam.  We suspect that the output mode profiles arise from a complex interplay between this nonlinear phase shift and self-induced transparency, which gives rise to a spatially-varying soft aperture effect that is dependent on the local intensity of the input mode.  Interference in the far-field then underlies the conversion from an input Gaussian beam to the non-Gaussian output mode.

\begin{figure}[t!]
\centering
\includegraphics*[width=\textwidth]{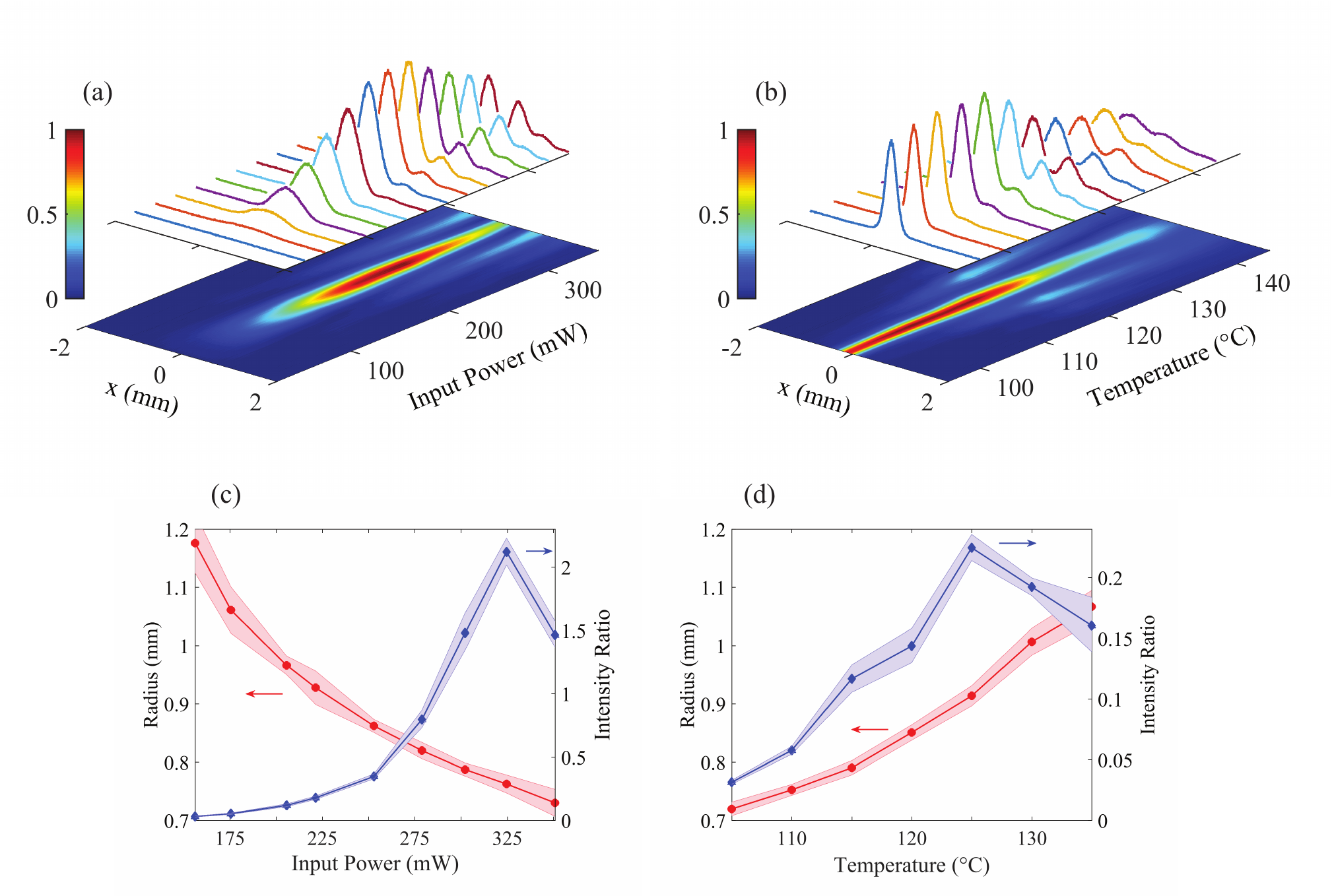}
\caption{ \textbf{Tunability of non-Gaussian modes obtained from atomic vapor by varying power and temperature.}  \textbf{(a)} Image cross-sections as a function of incident optical power, keeping the cell temperature fixed at $T = 125$ $^\circ$C. \textbf{(b)} Cross-sections as a function of cell temperature, with a constant input power of $P = 275$ mW.  \textbf{(c)} Radial positions (red) and intensities of the first ring relative to the central peak (blue) versus input optical power. $T = 125$ $^\circ$C.  \textbf{(d)} Radial positions (red) and relatives intensities of the first ring (blue) with varying cell temperature. $P = 250$ mW.  The shaded regions indicate uncertainties based on one standard deviation.}
\label{fig:slices}
\end{figure}

In Fig.~\ref{fig:slices} (a-b), we show two series of normalized intensity cross-sections obtained from single-shot images for various levels of input power and cell temperature.  Smoothed two-dimensional intensity projections of the measurements are displayed below.  We find that the conversion into the non-Gaussian pattern is enhanced for increasing levels of both optical power and temperature: rings with appreciable signal-to-noise ratios are generally observed at optical powers above $P \sim 200$ mW and temperatures above $T ~\sim 125 ^\circ$ C. These results agree with previous work~\cite{Zhang2015}, where the intensity-dependent phase shift and temperature-dependent atomic density of the nonlinear medium were considered.

We use numerical fitting to extract the radial positions and intensities (relative to the central peak) of the first ring in the output patterns obtained at various powers and temperatures. We observe that the atomic vapor acts as a nonlinear lens, as shown in Fig.~\ref{fig:slices} (c-d).  With increasing optical power, the atomic vapor focuses the converted mode and reduces the size of the ring in the far-field.  The relative intensity of the first ring increases even more dramatically with increasing power, reaching values greater than unity for input powers above 300 mW.  The maximum value of $P = 350$ mW used throughout the measurements was chosen based on the maximum allowed coupling power for a single-mode fiber.   Conversely, we see that increasing the cell temperature leads to beam defocusing.  The complex interplay of nonlinear effects is exemplified by the fact that increasing the cell temperature also increases the atomic density of the nonlinear medium, and we see that mode conversion is enhanced at higher temperatures, as evidenced by larger field intensity in the first ring at higher temperature.  In our experiments, mode conversion is optimal at a temperature of about T $\sim$ 125 $^\circ$C, as shown in Fig.~\ref{fig:slices} (d).  Above this temperature, the medium becomes less transparent and linear absorption limits the conversion.   

Given that the atomic vapor creates a nonlinear lensing effect, the focusing (or defocusing) strength of the output mode depends on the position of the vapor cell relative to the focus of the input beam.  In each experiment, effort was taken to position the cell at the center of the input beam's focus.  This was found to reliably give results such as those shown in Figs.~\ref{fig:slices} (a-b).  For the measurements in Figs.~\ref{fig:slices} (c-d), however, we found that the position of the cell could be adjusted to optimize the intensity of the ring.  In this case, the position was set to optimize the ring intensity, which reached a maximum at an input power of about $P ~\sim 325$ mW.  Accordingly, the results in Figs.~\ref{fig:slices} (c-d) differ from those in Figs.~\ref{fig:slices} (a-b).  Nonetheless, it was found to be qualitatively true that increasing the power and temperature would result in nonlinear focusing and defocusing of the mode, respectively.

Lastly, as expected, we have found that the mode shapes can be tuned by varying the laser frequency.  These results are shown in the supplementary information.  For this reason, the laser frequency was locked to a fixed detuning from the resonance (see Methods), to prevent fluctuations and/or drift in the laser frequency from altering the shape of the optical modes. 

\begin{figure}[ht!]
\centering
\includegraphics*[width=\textwidth]{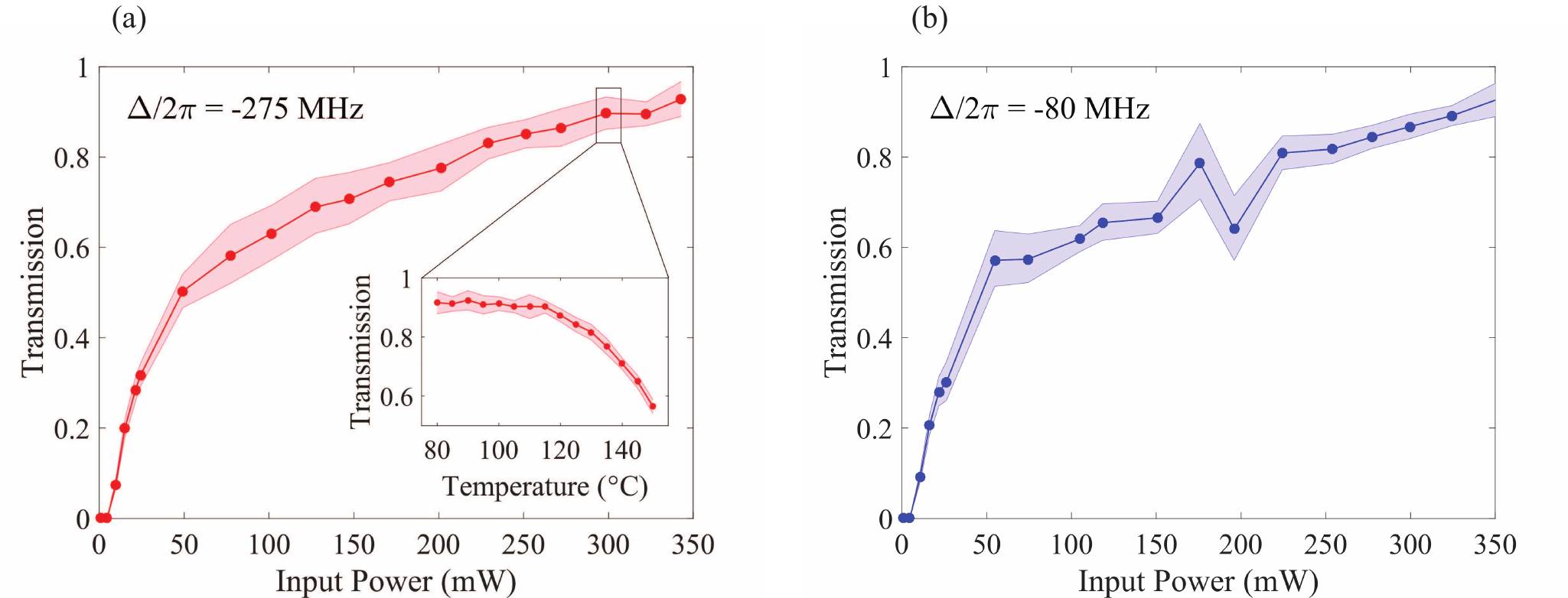}
\caption{\textbf{Self-induced transparency in atomic vapor}. \textbf{(a)}  Measured optical transmission through rubidium-85 vapor as a function of input optical power with a fixed cell temperature of $T = 125$ $^\circ$C and with the laser detuned 275 MHz to the red side of the D1 line.  The inset shows the corresponding optical transmission when varying cell temperature and keeping the input power fixed at $P =  300$ mW.  \textbf{(b)}.  Optical transmission as a function of input optical power for $T = 125$ $^\circ$C and a laser detuning of -80 MHz. The shaded regions indicate uncertainties based on standard deviations.}
\label{fig:transparency}
\end{figure}

These observations most likely stem from a combination of numerous competing self-actions effects.  The collection of these effects, however, can be thought of as an complex aperture which is tuned by the interactions between light and matter.  Among these interactions, we find that one of the dominant effects is a self-induced transparency of the medium~\cite{McCall1967, McCall1969, McClelland1986}, which effectively behaves as a soft-aperture for the input beam~\cite{LeBerre1984}.  In Fig.~\ref{fig:transparency} (a), we show the measured transmission (i.e., the ratio of the output optical power to input power) through the atomic vapor while the cell temperature is kept fixed at T$ = 125$ $^\circ$C.  At low ($\sim$ mW) power levels, optical losses resulting from atomic absorption strongly attenuate the beam.  As the power is increased, self-transparency occurs and the transmission increases rapidly, reaching 50$\%$ at an input power of $P \sim 50$ mW.  For higher powers, the rate of increase in transmission slows down and approaches unity.  A corresponding temperature measurement is shown in the inset of Fig.~\ref{fig:transparency} (a), for an input power of $P = 300$\,mW.  Here, the transmission falls off predictably at high temperatures.  We confirmed the transparency effect for frequency detunings of -275 MHz (Fig.~\ref{fig:transparency} (a) and -80 MHz (Fig.~\ref{fig:transparency} (b), the predominant frequencies used throughout this work.  The effect is typified by the fact that, at a given temperature, the transmission can be boosted simply by increasing the intensity of the optical beam.  

\subsection*{Reconstruction of non-Gaussians beams following an obstruction}
\begin{figure}[ht!]
\centering
\includegraphics*[width=\textwidth]{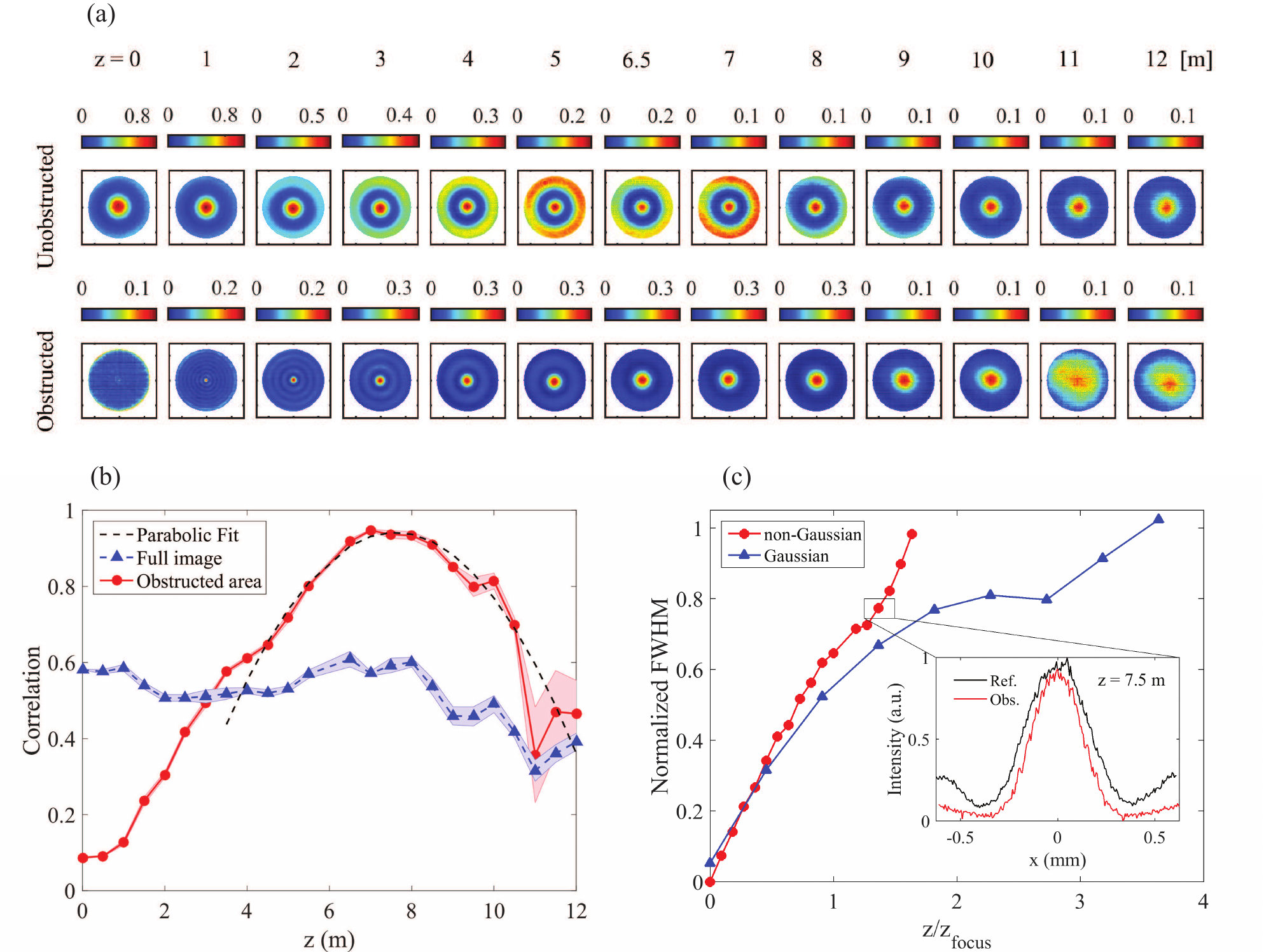}
\caption{\textbf{Self-reconstruction of a non-Gaussian mode generated with atomic vapor.}  \textbf{(a)} Images of unobstructed (top) and obstructed (bottom) modes at various positions along the propagation coordinate.  The size of each image is 1.5 mm $\times$ 1.5 mm, and the images correspond to the region over which the obstructed area correlation is calculated.  \textbf{(b)}  Calculated two-dimensional correlation between the (unobstructed) reference at $z = 0$ m and the obstructed modes at various positions for the full image (blue) and the obstruction area only (red).  In the latter case, a parabolic fit reveals an optimal correlation at $z \sim 7.5$ m.  The shaded regions indicate uncertainties based on one standard deviation. \textbf{(c)} Full width half maxima of central peaks normalized to their widths at $z = 0$ m, for a BG-like mode with a focal position $z_{\textrm{focus}} = 5.5$ m and a Gaussian mode with $z_{\textrm{focus}} = 1.1$ m.  In both cases, approximately 20\% of the light is blocked.  The inset shows cross-sections of the reference and obstructed BG-like modes at the optimal reconstruction distance of $z = 7.5$ m.}
\label{fig:reconstruction}
\end{figure}

The ability of an optical beam to regenerate itself following a disturbance is, in itself, a particularly interesting property given the diffractive nature of light.  Here we show that our non-Gaussian pattern can reconstruct itself after encountering a circular obstruction placed at the center of the optical mode. In Fig.~\ref{fig:reconstruction} (a), we show a series of normalized images which zoom in on the obstruction area and illustrate the reconstruction of the central portion of the mode as it travels along the propagation direction.  Images of the unobstructed and obstructed modes are shown in the top and bottom of Fig.~\ref{fig:reconstruction} (a), respectively.  At $z = 0$ m, the (removable) circular block attentuates approximately 20$\%$ of the light, and the remaining light increasingly regenerates itself for $z > 0$ m.  (A series of full images is shown in the Supplementary information.)  We quantify the reconstruction by calculating the two-dimensional correlation between the images of the (unobstructed) reference at $z = 0$ m and the propagating obstructed beam (see Supplementary information).  The result is shown in Fig.~\ref{fig:reconstruction} (b), where we have carried out two calculations: one taken over the entire area of the image, and a second using only the obstructed area.  Within the obstruction area, mode regeneration increases with $z$, reaching a maximum correlation of $\sim$\,94\% at $z \sim 7.5$ m.  If the calculation is taken over the entire image, maximum correlation is observed at $z \sim 8$ m, but as a whole, the correlation decreases with $z$.  Given that the modes resemble truncated (after the first- and second-order rings) BG modes, reconstruction, albeit imperfect, after a finite propagation distance is expected.


Next, we compare the reconstruction of our BG-like mode with that of a Gaussian mode.  Studying the propagation of the unobstructed modes in Fig.~\ref{fig:reconstruction} (a), one sees that the center of the mode reaches a focus around $z_{\textrm{focus}} \sim 5.5$ m.  Therefore, we consider the reconstruction of a focused Gaussian mode which has also been attenuated by approximately 20\% via a central obstruction.  We employ Gaussian fitting to extract the full width half maxima (FWHM) of the central peaks from the reconstructing modes in both cases.  In Fig.~\ref{fig:reconstruction} (c), we show the FWHM from single shot images normalized to their widths at $z = 0$ m, as a function of the normalized distance $z / z_{\textrm{focus}}$.  In both cases, mode regeneration is associated with an increase in the FWHM, which slows after the focus and eventually increases again due to diffraction.  Importantly, the normalized FWHM of the non-Gaussian mode reaches unity prior to the Gaussian, by more than a factor of two in normalized distance.  The point of maximum image correlation for the non-Gaussian case ($z \sim 7.5$ m) corresponds to the point $z / z_{\textrm{focus}} = 1.4$ in Fig.~\ref{fig:reconstruction} (c).  At this point, the normalized FWHM is $\sim$ 0.8 for the non-Gaussian mode (see the inset of Fig.~\ref{fig:reconstruction} (c)), whereas it is only $\sim$ 0.7 for the Gaussian mode.

\section*{Discussion}

In this manuscript we have experimentally demonstrated that atomic vapor may be used to generate tunable self-reconstructing optical beams.  While the complete dynamics of near-resonant light interacting with multi-level atomic systems is reasonably complex, these results suggest that nonlinear phase shifts and self-induced transparency are dominant effects, producing an effective soft aperturing of the incident optical mode, which leads to the conversion into a truncated BG mode.  Although this truncation is expected to worsen the mode's self-healing ability when compared to an ideal BG mode, we nonetheless observe image correlations of up to 94\% in the central portion of the recovered mode profile and up to 61\% across the entire image.  The flexible approach taken here is advantageously robust to high levels of optical power, and thus allows one to control the output mode profile by varying the optical power, as well as the temperature of the atomic vapor and the laser frequency.  In this manner, the reconstruction of the mode after interacting with an obstruction can be optimized, which may see applications in future experiments involving the recovery of information in the spatial profile of the modes.  We expect that such optimization could further improve the correlations reported here.

\section*{Methods}
\label{sec:methods}
Locking to the atomic resonances of rubidium is achieved with a polarization spectroscopy setup as described in Refs.~\cite{Wieman1976, Pearman2002, Ratnapala2004}.  Images are recorded using a beam profiler (Edmund Optics 89-308) with the laser locked at a fixed detuning $\Delta / 2 \pi = -80$ MHz from the red side of the D1 line of rubidium-85.  In Fig.~\ref{fig:transparency}, data is also shown for a second detuning of -275 MHz.  In each measurement, a series of ten images acquired over a span of approximately 1 s are recorded, and the data and error bars in Figs.~\ref{fig:slices} and ~\ref{fig:reconstruction} represent mean values and standard deviations, respectively, based on these ten images.  Every image is acquired using an integration time of 10 ms, and all of the actual images shown in this manuscript are single shots.

In the self-healing experiments, the obstruction is made by coating the surface of a circular object ($\sim$ 3 mm in diameter) with acrylic paint, and then contacting the object with a 170 $\mu$m thick, 22 mm $\times$ 22 mm precision microscope cover slip (Thorlabs CG15CH).  The obstruction is placed on a flip-mount so that both the reconstructing and the unobstructed propagating modes may be imaged. In comparing the reconstruction with that of a Gaussian beam, the obstruction is made using the same approach, but with the diameter chosen so that the object blocks $\sim$ 20\% of the light, as in the non-Gaussian mode case. The FWHM in Fig.~\ref{fig:reconstruction} are found via Gaussian fitting, and the focal positions $z_{\textrm{focus}}$ are taken to be the positions where the unobstructed FWHM of the central peaks are minimal.   


\section*{Acknowledgments}

We acknowledge funding from the Louisiana State Board of Regents and Northrop Grumman \emph{NG - NEXT}.

\section*{Author Contributions}

J.D.S performed the experiments, with contributions from K.N.D., E.M.K. and C. R.  J.D.S performed the data analysis, with contributions from O.D., and J.D.S and R.T.G wrote the manuscript.  R.T.G. conceived the project.

\section*{Additional Information}

\noindent \textbf{Supplementary information} accompanies this paper at \textbf{INSERT URL}.

\noindent \textbf{Competing financial interests:} The authors declare no competing financial interests. 

\end{document}